\documentclass[aps,prb,twocolumn,superscriptaddress,showpacs,floatfix]{revtex4}

\usepackage{tipa}
\usepackage{bbding}
\usepackage{txfonts}
\usepackage{amssymb}
\usepackage{graphicx}

\begin{document}

\title{Coherent population transfer between weakly-coupled states in a ladder-type superconducting qutrit}

\author{H. K. Xu}
\affiliation{Beijing National Laboratory for Condensed Matter Physics, Institute of
Physics, Chinese Academy of Sciences, Beijing 100190, China}
\author{W. Y. Liu}
\affiliation{Beijing National Laboratory for Condensed Matter Physics, Institute of
Physics, Chinese Academy of Sciences, Beijing 100190, China}
\author{G. M. Xue}
\affiliation{Beijing National Laboratory for Condensed Matter Physics, Institute of
Physics, Chinese Academy of Sciences, Beijing 100190, China}
\author{F. F. Su}
\affiliation{Beijing National Laboratory for Condensed Matter Physics, Institute of
Physics, Chinese Academy of Sciences, Beijing 100190, China}
\author{H. Deng}
\affiliation{Beijing National Laboratory for Condensed Matter Physics, Institute of
Physics, Chinese Academy of Sciences, Beijing 100190, China}
\author{Ye~Tian}
\affiliation{Beijing National Laboratory for Condensed Matter Physics, Institute of
Physics, Chinese Academy of Sciences, Beijing 100190, China}
\author{\mbox{D. N. Zheng}}
\affiliation{Beijing National Laboratory for Condensed Matter Physics, Institute of
Physics, Chinese Academy of Sciences, Beijing 100190, China}
\author{\mbox{Siyuan Han}}
\affiliation{Department of Physics and Astronomy, University of Kansas,Lawrence, KS
66045, USA}
\affiliation{Beijing National Laboratory for Condensed Matter Physics, Institute of
Physics, Chinese Academy of Sciences, Beijing 100190, China}
\author{Y. P. Zhong}
\affiliation{Department of Physics, Zhejiang University, Hangzhou 310027, China}
\author{H. Wang}
\affiliation{Department of Physics, Zhejiang University, Hangzhou 310027, China}
\author{Yu-Xi Liu}
\affiliation{Institute of Microelectronics, Tsinghua University, Beijing 100084, China}
\author{S. P. Zhao}
\affiliation{Beijing National Laboratory for Condensed Matter Physics, Institute of
Physics, Chinese Academy of Sciences, Beijing 100190, China}

\begin{abstract}
Stimulated Raman adiabatic passage (STIRAP) offers significant
advantages for coherent population transfer between un- or
weakly-coupled states and has the potential of realizing efficient
quantum gate, qubit entanglement, and quantum information transfer.
Here we report on the realization of STIRAP in a superconducting
phase qutrit - a ladder-type system in which the ground state
population is coherently transferred to the second-excited state via
the dark state subspace. The result agrees well with the numerical
simulation of the master equation, which further demonstrates that
with the state-of-the-art superconducting qutrits the transfer
efficiency readily exceeds $99\%$ while keeping the population in
the first-excited state below $1\%$. We show that population
transfer via STIRAP is significantly more robust against variations
of the experimental parameters compared to that via the conventional
resonant $\pi$ pulse method. Our work opens up a new venue for
exploring STIRAP for quantum information processing using the
superconducting artificial atoms.
\end{abstract}

\maketitle





Stimulated Raman adiabatic passage (STIRAP), which combines the
processes of stimulated Raman scattering and adiabatic passage, is a
powerful tool used for coherent population transfer (CPT) between
un- or weakly-coupled quantum states \cite{ber98,sho11}. It has been
recognized as an important technique in quantum computing and
circuit QED involving superconducting qubits
\cite{zho02,kis02,kis04,yan03,yan04,zho04,liu05,wei08,sie09,fal13}.
For example, qubit rotations can be realized via STIRAP with the two
computational states plus an auxiliary state forming a three-level
$\Lambda $ configuration \cite{zho02,kis02}. A scheme for generating
arbitrary rotation and entanglement in the three-level $\Lambda
$-type flux qutrits is also proposed \cite{kis04}, and the
experimental feasibility of realizing quantum information transfer
and entanglement between qubits inside microwave cavities has been
discussed \cite{yan03,yan04}. Unlike the conventional resonant $\pi$
pulse method STIRAP is known to be much more robust against
variations in experimental parameters, such as the frequency,
amplitude, and interaction time of microwave fields, and the
environmental noise \cite{kis02,kis04,wei08,sie09}.

Recently, multi-level systems (qutrits or qudits) have found
important applications in speeding up quantum gates \cite{lan09},
realizing quantum algorithms \cite{dic09}, simulating quantum
systems consisting of spins greater than $1/2$ \cite{nee09},
implementing full quantum-state tomography \cite{the02,bia10,sha13},
and testing quantum contextuality \cite{cab12}. Unlike the highly
anharmonic $\Lambda $-type flux qutrits the phase and transmon
qutrits have the ladder-type ($\Xi $-type) three-level configuration
which is weakly anharmonic. The dipole coupling between the ground
state $|0\rangle $ and the second-excited state $|2\rangle $ in the
phase qutrit is much weaker than those between the first-excited
state $|1\rangle $ and the $|0\rangle $ state or the $|2\rangle $
state. In the case of the transmon qutrit the coupling is simply
zero. This unique property makes it difficult to transfer population
from $|0\rangle $ to $|2\rangle $ directly using a single $\pi $
pulse tuned to their level spacing $\omega _{20}$. The usual
solution is to use the high-power resonant two-photon process or to
apply two successive $\pi $ pulses transferring the population first
from $|0\rangle $ to $|1\rangle $ and then from $|1\rangle $ to
$|2\rangle $ \cite{bia10,sha13}. These methods often lead to a
significant population in the middle level $|1\rangle $ resulting in
energy relaxation which degrades the transfer process. In contrast,
STIRAP transfers the qutrit population directly from $|0\rangle $ to
$|2\rangle $ via the dark state subspace without occupying the
middle level $|1\rangle $.

In this work, we report on the realization of STIRAP in a $\Xi$-type
superconducting phase qutrit \cite{sim04}. As shown schematically in
Fig.~1a the qutrit has a loop inductance $L$ and a Josephson
junction with critical current $I_{c}$ and capacitance $C$. \ The
potential energy and quantized levels $|0\rangle $, $|1\rangle $,
and $|2\rangle $ of the qutrit are illustrated in Fig.~1b in which
the frequencies $\omega _{p,s}$ of the pump and Stokes fields and
their strength $\Omega _{p,s}$ (Rabi frequencies) are also
indicated. Applying the rotating-wave approximation (RWA) in the
double-rotating frame the Hamiltonian of the system can be written
as \cite{sil09,li11}:
\begin{equation}
H=\left[
\begin{array}{ccc}
0 & g_{p}+g_{s}e^{-i\delta t} & 0 \\
g_{p}+g_{s}e^{i\delta t} & \Delta _{p} & \lambda (g_{p}e^{i\delta
t}+g_{s})
\\
0 & \lambda (g_{p}e^{-i\delta t}+g_{s}) & \Delta _{p}+\Delta _{s}
\end{array}
\right] ,  \label{H_RWA}
\end{equation}
where the Planck constant $\hbar $ is set to unity, $\delta =\omega
_{p}-\omega _{s},$ $\Delta _{p}=\omega _{10}-\omega _{p}$ and
$\Delta _{s}=\omega _{21}-\omega _{s}$ are various detunings,
$g_{p,s}$ are the qutrit-microwave couplings proportional to the
amplitude of the pump and Stokes fields respectively, and $\lambda
\approx \sqrt{2}$ for the $\Xi $-type configuration with weak
anharmonicity. \ For $\delta \gg \Omega _{p,s}$ the fast-oscillating
terms in equation~(1) averages out to zero so the Hamiltonian
becomes
\begin{equation}
H'=\left[
\begin{array}{ccc}
0 & \Omega _{p}/2 & 0 \\
~~\Omega _{p}/2~~ & ~~\Delta _{p}~~ & \Omega _{s}/2 \\
0 & \Omega _{s}/2 & \Delta _{p}+\Delta _{s}%
\end{array}
\right] ,  \label{H_qutrit}
\end{equation}
in which $\Omega _{p}=2g_{p}$ and $\Omega _{s}=2\lambda g_{s}$. For
the phase qutrit used here we have $\lambda \simeq 1.45$. \
Equation~(\ref{H_qutrit}) is the well-known RWA Raman Hamiltonian
\cite{ber98,sho11}. In particular, when the system satisfies the
two-photon resonant condition:
\begin{equation}
\Delta _{p}+\Delta _{s}=0,  \label{2-photon}
\end{equation}
it has an eigenstate $|D\rangle =\cos \Theta |0\rangle -\sin \Theta
|2\rangle $, called the dark state, which corresponds to the
eigenvalue of $\epsilon =0$. Here $\tan \Theta(t) =\Omega
_{p}(t)/\Omega _{s}(t) $. CPT from the ground state $|0\rangle $ to
the second-excited state $|2\rangle $ without populating the
first-excited state $|1\rangle $ can therefore be realized via
STIRAP by initializing the qutrit in the ground state $|0\rangle $
\cite{li11,nov13} and then slowly increasing the ratio $\Omega
_{p}(t)/\Omega _{s}(t)$ to infinity as long as the following
conditions \cite{ber98,sho11,scu97,vas09}
\begin{equation}
\delta \gg \Omega _{p,s},~\text{and }\int_{-\infty }^{\infty
}\sqrt{\Omega _{p}^{2}(t)+\Omega _{s}^{2}(t)}~dt~>10\pi
\label{condition}
\end{equation}
are satisfied so that the qutrit will stay in the dark state
subspace spanned by \{$|0\rangle ,|2\rangle $\}. The first condition
is required to reduce equation~(\ref{H_RWA}) to
equation~(\ref{H_qutrit}) leading to the existence of the dark state
solution while the second ensures adiabatic state following.\\

\begin{figure}[t]
\includegraphics[width=0.35\textwidth,angle=-90]{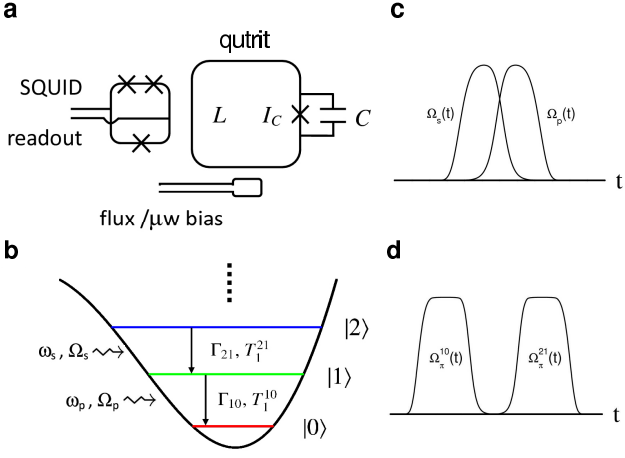}
\caption{{\bf Superconducting phase qutrit and measurement pulse
sequences.} ({\bf a}) Schematic rf-SQUID type phase qutrit with
Josephson critical current $I_c$, shunt capacitance $C$, and loop
inductance $L$. ({\bf b}) Three bottom energy levels
$\mid$0$\rangle$, $\mid$1$\rangle$, and $\mid$2$\rangle$ of the
qutrit with related symbols indicated. Subscripts $p$ and $s$ refer
to the pump and Stokes tones, respectively. ({\bf c})
Counterintuitive pulse sequence with $\Omega_s$ preceding $\Omega_p$
for coherent population transfer from $\mid$0$\rangle$ to
$\mid$2$\rangle$ without involving $\mid$1$\rangle$. ({\bf d})
Conventional resonant $\pi $ pulse sequence for successive
$|0\rangle \rightarrow |1\rangle $ $\rightarrow $ $|2\rangle $
population transfers.} \label{Fig-1}
\end{figure}

\noindent {\textbf{\textsf{Results}}}

\noindent {\bf Sample parameters and measurements.} The sample used
in this work is an aluminum phase qutrit \cite{sim04}, which is
cooled down to $T\approx 10$ mK in an Oxford cryogen-free dilution
refrigerator. The qutrit control and measurement circuit includes
various filtering, attenuation, and amplification \cite{tia12}. For
the present experiment, we bias the rf-SQUID to have six energy
levels in the upper potential well and use the lowest three levels
as the qutrit states. The relevant transition frequencies are
$f_{10}$ $=\omega _{10}/2\pi =5.555$ GHz and $f_{21}=\omega
_{21}/2\pi =5.393$ GHz, and the relative anharmonicity is $\alpha
=(f_{10}-f_{21})/f_{10}\approx 2.9\%$. The measured energy
relaxation times are $T_{1}^{10}=1/\Gamma _{10}=353$ ns and
$T_{1}^{21}=1/\Gamma _{21}=196$ ns, respectively, while the
dephasing time determined from Ramsey interference experiment is
$T_{\varphi }^{10}=124$ ns. To realize STIRAP, a pair of bell-shaped
counterintuitive microwave pulses with the Stokes pulse preceding
the pump pulse, as illustrated in Fig.~1c, are used. The pulses are
defined by $\Omega _{s}(t)=\Omega _{0}F(t)\cos [\pi f(t)/2]$ and
$\Omega _{p}(t)=\Omega
_{0}F(t)\sin [\pi f(t)/2]$ with $F(t)=e^{-(t/2T_{d})^{6}}$ and $%
f(t)=1/(1+e^{-4t/T_{d}})$ \cite{sho11,vas09}. The pulse width (FWHM)
is approximately $2T_{d}$, and its height is $0.967\Omega _{0},$ and
$\Omega _{s}(0)=\Omega _{p}(0)$.

\begin{figure}[t]
\includegraphics[width=0.42\textwidth]{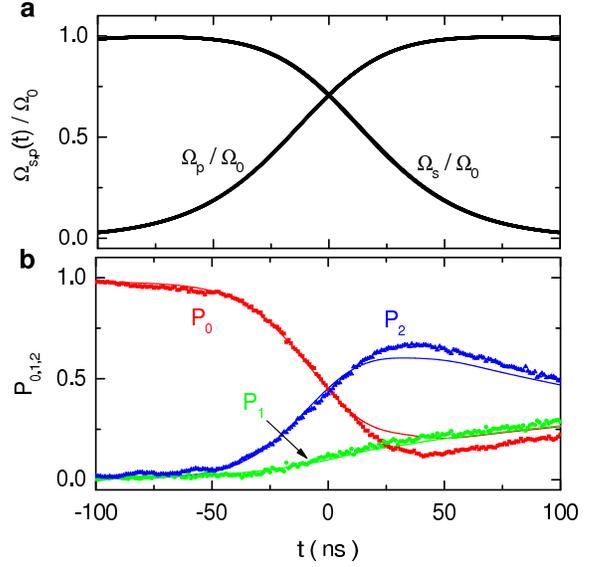}
\caption{{\bf Coherent population transfer via STIRAP in the
superconducting phase qutrit.} ({\bf a}) Stokes and pump microwave
pulses $\Omega_s(t)$ and $\Omega_p(t)$ shown in the overlapped
region with the experimental parameters $\Omega_{0}/2\pi$ =
42.8$\pm$1.0 MHz and $T_d$ = 100 ns. ({\bf b}) Level populations
$P_0$, $P_1 $, and $P_2$ versus time in the case of $\Delta_p$ =
$\Delta_s$ = 0. A maximum experimental value of $P_2$ = 67$\%$,
limited by the low decoherence times of the device, is reached.
Experimental and calculated results are shown as symbols and lines,
respectively.} \label{Fig-2}
\end{figure}

\noindent {\bf Coherent population transfer.} Figure~2a shows the
two microwave pulses defined by $\Omega _{0}/2\pi $ $=42.8\pm 1.0$
MHz and $T_{d}$ $=100$ ns in their overlapping region. As $t$
increases, $\Omega _{s}(t)$ and $\Omega _{p}(t)$ start to increase
and decrease across $t=0$ at which they are equal. The
experimentally measured populations $P_{0}$, $P_{1}$, and $P_{2}$
versus time produced by this counterintuitive pulse sequence in the
resonant case $\Delta _{p}=\Delta _{s}=0$ are plotted in Fig.~2b as
symbols. We observe that as time evolves across $t=0$ the population
$P_{2}$ ($P_{0}$) increases (decreases) rapidly, signifying the
occurrence of STIRAP via the dark state of the superconducting
qutrit system. The experimentally achieved maximum $P_{2}$, or the
population transfer efficiency, is about $67\%$ for the present
sample under the resonant condition. Notice that in the entire
region of $t\in \lbrack -100$, $100]$ ns, all of the characteristic
features of the experimental data, in particular (i) $P_{1}$
remaining significantly lower than $P_{2}$ for $t>-50$ ns, (ii) the
decrease (increase) of $P_{2}$ ($P_{0}$) after reaching the maximum
(minimum) as well as the gradual rising of $P_{1}$, are reproduced
well by the numerical simulation (solid lines). The simulated
temporal profiles of the populations $P_{0},$ $P_{1}$, and $P_{2}$
are obtained by solving the master equation $\dot{\rho}=-(i/\hbar
)[H,\rho ]+L(\rho ),$ using the measured qutrit parameters, where
$L(\rho )$ is the Liouvillean containing the relaxation and
dephasing processes \cite{li11}. The numerical result also confirms
that feature (ii) is due primarily to energy relaxation, while the
maximum $P_{2}$ itself is mainly limited by dephasing. Hence by
increasing the qutrit decoherence times the undesirable decrease of
$P_{2}$ and the rise of $P_{1}$ and $P_{0}$ in the relevant time
scale can be suppressed and a higher $P_{2}$ can be achieved (see
below).

In our experiment the conditions imposed by
equation~(\ref{condition}) are satisfied: $\delta /2\pi $ in the
resonant case $\Delta _{p}$ $=\Delta _{s}=0$ is $f_{10}-f_{21}=162$
MHz, which is approximately four times that of $\Omega _{0}/2\pi $,
and it is easy to verify that the integrated pulse area
$\int_{-\infty }^{\infty } \sqrt{\Omega _{p}^{2}(t)+\Omega
_{s}^{2}(t)}~dt$ $\approx $ $32\pi $ is greater than $10\pi $. In
addition to reducing decoherence the efficiency of the demonstrated
STIRAP process can be improved by increasing the relatively small
anharmonicity parameter $\alpha \approx 2.9\%$ of the present sample
up to, say, $10\%$ by optimizing device parameters of the $\Xi
$-type phase and transmon (or Xmon) qutrits \cite{whi14,hoi13}.
According to equation~(\ref{condition}) greater anharmonicity allows
the use of larger $\Omega _{0}$ which would proportionally reduce
the duration of the pump and Stokes pulses when the pulse area is
kept unchanged to satisfy the adiabatic condition. Shorter pulses
also reduce the negative effect of decoherence on the transfer
efficiency.

\begin{figure}[t]
\includegraphics[width=0.42\textwidth]{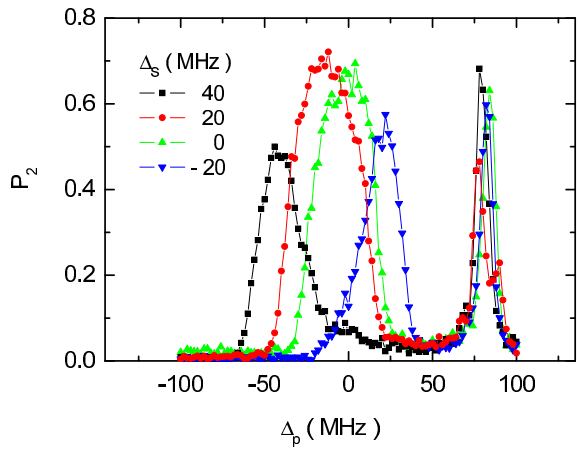}
\caption{{\bf Level population $P_2$ versus pump microwave detuning
$\Delta_p$ for four Stokes tone detunings.} Bright resonances can be
seen as the resonant condition equation~(3) is satisfied (left-side
peaks in each curves). A maximum value of $P_2$ = 72$\%$ is reached
at the microwave detunings $\Delta_p$ = -$\Delta_s$ = -20 MHz. The
right-side peaks result from the two-photon process of the single
pump microwave tone.} \label{Fig-3}
\end{figure}

The STIRAP process is often identified in either the time domain or
the frequency domain \cite{ber98,sho11}. The latter is based on
equation~(\ref{2-photon}) which specifies the two-photon resonance
condition. In Fig.~3, we show the level $|2\rangle $ population
$P_{2}$ versus the pump detuning $\Delta _{p}$ for four Stokes tone
detunings of $\Delta _{s} = 40, 20, 0$, and -$20$ MHz, respectively.
Bright resonance appears as the left-side peak in each curve when
the two-photon resonant condition equation~(\ref{2-photon}) is met.
It is interesting to see that the maximum value of $P_{2}=72\%$ is
reached at the microwave detunings of $\Delta _{p}=-\Delta _{s} =
-20$ MHz, which is higher than the value achieved in the resonant
case of $\Delta _{p} = \Delta _{s}=0$ shown in Fig.~2b. In Fig.~3,
the right-side peak in each curve is originated from the two-photon
process excited by the single pump microwave tone. Compared to the
left-side peaks, although the peak heights are comparable, they are
much narrower, indicating that in practice it is less controllable
using the two-photon process to perform coherent population transfer
from state $|0\rangle $ to state $|2\rangle $.

\noindent {\bf Efficiency and robustness.} The above results
demonstrate clearly CPT from the ground state $|0\rangle $ to the
second-excited state $|2\rangle $ via STIRAP in the $\Xi $-type
superconducting qutrit. Compared to the usual high-power two-photon
process or two non-overlapping successive resonant $\pi $ pulse
excitations shown in Fig.~1d, which involve significant undesired
population in the middle level $|1\rangle $ and require precise
single photon resonance and pulse area \cite{wei08,bia10}, CPT via
STIRAP demonstrates just the opposite. First, in principle CPT
between $|0\rangle $ and $|2\rangle $ can be accomplished without
occupying the lossy middle level $|1\rangle $. More importantly, the
process is much more robust against variations in the frequency,
duration, and shape of the driving pulses \cite{ber98,sho11}. In
fact, in terms of the conditions equation~(\ref{2-photon}) and
equation~(\ref{condition}), we see from Fig.~3 that the single
photon resonance condition is greatly relaxed. Although $\Omega
_{p,s}$ are limited by the system anharmonicity, their values,
together with $T_{d}$, still have much room for variations while
maintaining the transfer efficiency.

\begin{figure}[t]
\includegraphics[width=0.48\textwidth]{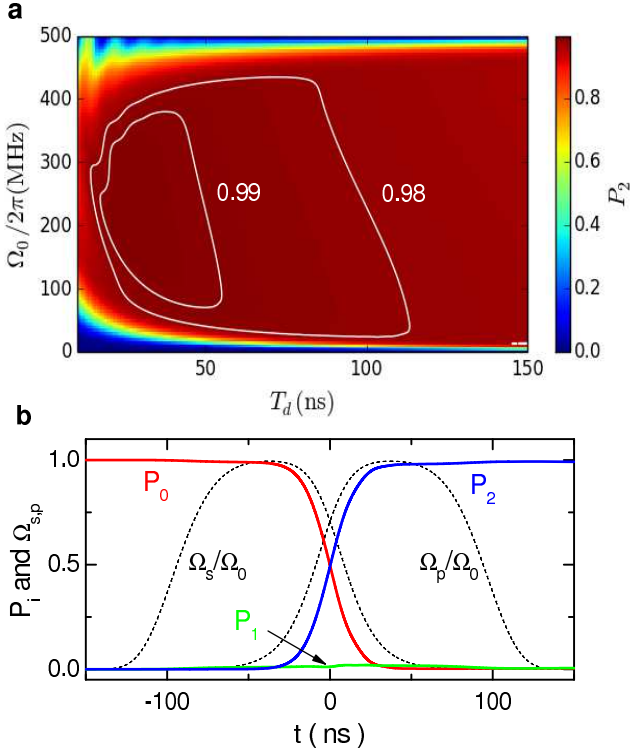}
\caption{{\bf Calculated results demonstrating the robustness of the
STIRAP process.} ({\bf a}) Results in the $\Omega_0$ versus $T_d$
plane with contours for population transfer efficiency above 99$\%$
and 98$\%$ indicated. The same sample parameters as those in Fig.~2b
are used in the calculation but the decoherence times are increased
to $T_1^{10}$ = 35.3 $\protect\mu$s, $T_1^{21}$ = 19.6 $\mu$s, and
$T_{\varphi}^{10}$ = 12.4 $\mu$s and the relative anharmonicity to
$\alpha = 8\%$. ({\bf b}) The temporal profiles of microwave pulses
(dashed lines) and level populations (solid lines) in the case of
$\Omega_0 /2\pi$ = 100 MHz and $T_d$ = 50 ns.} \label{Fig-4}
\end{figure}

In Fig.~4a, we show the calculated results in the $\Omega _{0}$
versus $T_{d}$ plane with contours indicating population transfer
efficiency above $99\%$ and $98\%$, respectively, using qutrits with
decoherence times of $T_{1}^{10}$ = 35.3 $\mu $s, $T_{1}^{21}$ =
19.6 $\mu $s, and $T_{\varphi }^{10}$ = 12.4 $\mu $s, and the
relative anharmonicity of $\alpha = 8\%$. The qutrit parameters in
the ranges are now attainable with transmon- \cite{pai11,hoi13},
Xmon- \cite{bar13}, flux- \cite{ste14}, and also possibly phase-
\cite{whi14} type devices. In Fig.~4b, we show the level populations
versus time (solid lines) for $\Omega_0 /2\pi$ = 100 MHz and $T_d$ =
50 ns, in which a nearly complete transfer above 99$\%$ from level
$|0\rangle $ to level $|2\rangle $ is accomplished with the
population in middle level $|1\rangle $ kept below 1$\%$. These
results indicate that the transfer efficiency of STIRAP is very
insensitive to $\Omega_0$, which is limited by systems
anharmonicity, and to $T_d$, which should be much smaller than the
decoherence time. The allowed variations of a few hundreds of MHz in
$\Omega_0$ and a few tens of ns in $T_d$ for keeping $P_2$ $>$
99$\%$, for example, are in sharp contrast with the case of simple
$\pi$ pulse excitations. In fact, using a $\pi$ pulse to transfer
population from the ground state $|0\rangle $ to the first-excited
state $|1\rangle $ alone, acceptable variations of the Rabi
frequency $\Omega$ and pulse width $T$ to keep the transfer
efficiency above 99$\%$ can be estimated from the pulse area
relation $\Omega \times T$ $\approx $ $(1\pm 0.06)\pi$. Thus if we
use $\Omega /2\pi$ = 50 MHz, the pulse width variation must be less
than $\pm$ 0.6 ns. More strict condition is required when successive
excitation from the first-excited state $|1\rangle $ to the
second-excited state $|2\rangle $ is considered. From these we see
that the extreme robustness of the STIRAP process is very
advantageous and should be useful in various applications such as
realizing efficient qubit rotation, entanglement, and quantum
information transfer in various superconducting qubit and qutrit systems.\\

\noindent {\textbf{\textsf{Discussion}}}

We have experimentally demonstrated coherent population transfer
between two weakly-coupled states, $|0\rangle $ and $|2\rangle $, of
a superconducting phase qutrit having the $\Xi $-type ladder
configuration via STIRAP. The qutrit had a small relative
anharmonicity of $2.9\%$, and moderate decoherence times of
$T_{\varphi }^{10}$ $=124$ ns, $T_{1}^{10}$ $=353$ ns, and
$T_{1}^{21}=196$ ns, respectively. We demonstrated that by applying
a pair of counterintuitive microwave pulses in which the Stokes tone
precedes the pump tone, coherent population transfer from $|0\rangle
$ to $|2\rangle $ with a $72\%$ efficiency can be achieved with a
much smaller population in the first-excited state $|1\rangle $.
Using the measured qutrit parameters, including decoherence times,
we simulated the STIRAP process by numerically solving the master
equation. The result agrees well with the experimental data. We
showed that by increasing the decoherence times of the qutrits to
the order of a few tens of microseconds, currently attainable
experimentally, the transfer efficiency can be increased to greater
than $99\%$ while keeping the population of the first-excited state
below $1\%$.

We have also shown that coherent population transfer via STIRAP is
much more robust against variations of the experimental parameters,
including the amplitude, detuning, and time duration of the
microwave fields, and the environmental noise over the conventional
methods such as using high-power two-photon excitation and two
resonant $\pi $ pulses tuned to $\omega _{10}$ and $\omega _{21}$,
respectively. Therefore STIRAP is advantageous for achieving robust
coherent population transfer in the ladder-type superconducting
artificial atoms that play increasingly important roles in various
fields ranging from fundamental physics to quantum information
processing. Furthermore, the method can be readily extended to the
$\Lambda $-type systems such as the superconducting flux qutrits, in
which the initial and target states locate in different potential
wells representing circulating currents in opposite directions. Our
work paves the way for further progress in
these directions.\\

\noindent {\textbf{\textsf{Methods}}}

\noindent{\bf Determination of level populations.} The qutrit level
populations $P_{i}(t)$ ($i=0,1,2$) at a given time $t$ are
determined using two carefully calibrated nanosecond-scale
measurement flux pulses A and B, which reduce the potential barrier
to two different levels so that tunneling probabilities $p^{A}$ and
$p^{B}$ in each case are measured. Pulse A leads to low tunneling
probability $p_{0}^{A}(\sim 5\%)$ for state $|0\rangle $, high
tunneling probability $p_{1}^{A}$ for state $|1\rangle $ and, of
course, even higher tunneling probability $p_{2}^{A}$ for state
$|2\rangle $ . Pulse B results in a slightly deeper potential well
than pulse A does so that $p_{0}^{B}\simeq 0$, $\ p_{1}^{B}\approx
5\%$, and a much larger $ p_{2}^{B}$ for state $|0\rangle $,
$|1\rangle ,$ and $|2\rangle $ respectively. Denoting the density
operator of the qutrit as $\rho$, we have
\begin{equation}
p^{A,B}=P_{0}p_{0}^{A,B}+P_{1}p_{1}^{A,B}+P_{2}p_{2}^{A,B}~,
\end{equation}
\noindent where $P_{i}=\rho _{ii}$, and $p_{i}^{A,B}$ can be found
from the experimentally determined tunneling probabilities $p_i$ of
the $i$th energy level for given amplitudes of pulses A and B
\cite{sha13}. Combining the normalization condition
\begin{equation}
{\text Tr}\rho = P_0+P_1+P_2 = 1~,
\end{equation}
\noindent we obtain
\begin{equation}
P_i=[(p_j^B-p_k^B)p^A+(p_k^A-p_j^A)p^B+p_j^Ap_k^B-p_k^Ap_j^B]~/D~,
\end{equation}
\noindent where $i$ = 0, 1, 2 with $\{ i,j,k \}$ in circulative
order like $\{ 0,1,2 \}$, $\{ 1,2,0 \}$, and $\{ 2,0,1 \}$, and $D$
is the determinant
\begin{eqnarray}
D=\left|
\begin{array}{ccc}
1 & 1 & 1 \\
p_0^A & p_1^A & p_2^A \\
p_0^B & p_1^B & p_2^B
\end{array}
\right|~.
\end{eqnarray}
\noindent Hence the level populations $P_0$, $P_1$,and $P_2$ can be
obtained by measuring $p^A$ and $p^B$.

\noindent {\bf Numerical simulations.} We numerically calculate the
level populations $P_0(t)$ = $\rho_{00}(t)$, $P_1(t)$ =
$\rho_{11}(t)$, and $P_2(t)$ = $\rho_{22}(t)$ at any given time by
solving the master equation
\begin{equation}
\dot{\rho}=-\frac{i}{\hbar}[H,\rho]+L(\rho)~,
\end{equation}
\noindent where $\rho$ is the system's 3$\times$3 density matrix,
$H$ is the Hamiltonian given by equation~(1), and $L(\rho)$ is the
Liouvillean containing various relaxation and dephasing processes.
Since experimentally the pump and Stokes microwaves are not
correlated, we introduce a phase difference $\phi$ between the two
microwaves in the actual calculations \cite{li12}. In this case, the
double-rotating reference frame is described by the operator
$U=|0\rangle \langle 0|+|1\rangle \langle 1|e^{i\omega_p
t}+|2\rangle \langle 2|e^{i(\omega_p t+ \omega_s t-\phi)}$, and the
rotating-wave approximation leads to a Hamiltonian in the following
form:
\begin{widetext}
\begin{equation}
H=\left[
\begin{array}{ccc}
0 & g_{p}+g_{s}e^{-i(\delta t - \phi)} & 0 \\
g_{p}+g_{s}e^{i(\delta t - \phi)} & \Delta _{p} & \lambda
[g_{p}e^{i(\delta t - \phi)}+g_{s}] \\
0 & \lambda [g_{p}e^{-i(\delta t - \phi)}+g_{s}] & \Delta
_{p}+\Delta _{s}~.
\end{array}
\right], \label{H_RWA1}
\end{equation}
\end{widetext}
\noindent where the Liouvillean operator in equation~(9) is given by
\cite{li11}:
\begin{widetext}
\begin{eqnarray}
L(\rho)=-\frac{1}{2}\left[
\begin{array}{ccc}
-2\Gamma_{10}\rho_{11} & (\Gamma_{10}+\gamma_{10}^{\varphi})\rho_{01} & (\Gamma_{21}+\gamma_{20}^{\varphi})\rho_{02} \\
(\Gamma_{10}+\gamma_{10}^{\varphi})\rho_{10} & 2\Gamma_{10}\rho_{11}-2\Gamma_{21}\rho_{22} & (\Gamma_{10}+\Gamma_{21}+\gamma_{21}^{\varphi})\rho_{12} \\
(\Gamma_{21}+\gamma_{20}^{\varphi})\rho_{20} &
(\Gamma_{10}+\Gamma_{21}+\gamma_{21}^{\varphi})\rho_{21} &
2\Gamma_{21}\rho_{22}
\end{array}
\right].
\end{eqnarray}
\end{widetext}
We use the parameters $\Gamma_{10}$ = 2.83$\times$10$^6$ sec$^{-1}$,
$\Gamma_{21}$ = 5.10$\times$10$^6$ sec$^{-1}$, and
$\gamma_{10}^{\varphi}$ = 8.06$\times$10$^6$ sec$^{-1}$ measured
directly from experiment, while from the measured dc Stark shift of
the two-photon spectral lines we estimate $\gamma_{20}^{\varphi}$
$\approx$ 2$\gamma_{10}^{\varphi}$ and $\gamma_{21}^{\varphi}$
$\approx$ $\gamma_{10}^{\varphi}$. In our calculations
$\rho(t,\phi)$ is obtained by solving equation~(9) using the
fourth-order Runge-Kutta method. Since the phase difference $\phi$
of the two microwaves in our experiment is random, we average the
result over $\phi$ and finally arrive at:
\begin{equation}
\rho(t)=\frac{1}{2\pi}\int_0^{2\pi}\rho(t,\phi)d\phi~.
\end{equation}

\noindent {\textbf{\textsf{Acknowledgements}}}

We thank J. M. Martinis (UCSB) for providing us with the sample used
in this work. This work was supported by the Ministry of Science and
Technology of China (Grant Nos. 2011CBA00106, 2014CB921202, and
2015CB921104) and the National Natural Science Foundation of China
(Grant Nos. 91321208 and 11161130519). S. Han acknowledges support
by the US NSF (PHY-1314861).\\

\noindent {\textbf{\textsf{Author contributions}}}

H.K.X, S.H., and S.P.Z. designed the experiment. H.K.X., W.Y.L.,
G.M.X., and F.F.S. performed the measurement and numerical
simulation. Y.T., H.D., D.N.Z., Y.P.Z., and H.W. contributed to the
experiment in the low-temperature achievement, sample mounting and
characterization. Y.X.L. provided theoretical support. S.P.Z. and
S.H. wrote the manuscript in cooperation with all the authors.\\

\noindent {\textbf{\textsf{Additional information}}}

\noindent {\bf Competing financial interests}: The authors declare
no competing financial interests.

\end{document}